\documentclass[10pt]{article}
\usepackage{spconf,amsmath,graphicx}
\usepackage{amsfonts,amssymb}
\usepackage{multirow}
\usepackage{diagbox}
\usepackage{booktabs}
\usepackage{colortbl}
\usepackage{subfigure}
\usepackage{epstopdf}

\providecommand{\zd}[1]{\textcolor{black}{#1}}

\title{Spoofing Speaker Verification Systems with Deep Multi-Speaker Text-to-Speech Synthesis}
\twoauthors
 {Mingrui Yuan\sthanks{M. Yuan performed the work during his visit at University of Rochester.}}
	{Tsinghua University\\
	Dept. of Electronic Engineering\\
	Beijing, 100084, P.R. China\\
	\href{mailto:yuanmingrui1997@gmail.com}{\texttt{yuanmingrui1997@gmail.com}}}
 {Zhiyao Duan\sthanks{Z. Duan thanks the National Science Foundation grant No. 1617107.}}
	{University of Rochester\\
	Dept. of Electrical and Computer Engineering\\
	Rochester, NY 14627, USA\\
	\href{mailto:zhiyao.duan@rochester.edu}{\texttt{zhiyao.duan@rochester.edu}}}
\begin{document}
%
\maketitle

\begin{abstract}
This paper proposes a deep multi-speaker text-to-speech (TTS) model for spoofing speaker verification (SV) systems. The proposed model employs one network to synthesize time-downsampled mel-spectrograms from text input and another network to convert them to linear-frequency spectrograms, which are further converted to the time domain using the Griffin-Lim algorithm. Both networks are trained separately under the generative adversarial networks (GAN) framework. Spoofing experiments on two state-of-the-art SV systems (i-vectors and Google's GE2E) show that the proposed system can successfully spoof these systems with a high success rate. Spoofing experiments on anti-spoofing systems (i.e., binary classifiers for discriminating real and synthetic speech) also show a high spoof success rate when such anti-spoofing systems' structures are exposed to the proposed TTS system.

\end{abstract}
\section{Introduction}
\label{sec:intro}
Speaker verification (SV) is to verify whether a claim that an utterance belongs to a speaker is true or not. It is a widely used biometric and has been under research for decades. 
One widely used traditional method uses Gaussian mixture models (GMM) with a universal background model (UBM) \cite{reynolds2000speaker}. Later, i-vectors method was proposed, which maps high-dimensional statistics from the UBM into a low-dimensional representation \cite{dehak2010front}. Recently, deep learning methods have shown significant advances on verification accuracy over traditional methods \cite{snyder2017deep,snyder2018x,DBLP:journals/corr/LiMJLZLCKZ17,wan2018generalized}.

Similar to verification systems using other biometrics, SV systems face the problem of fake identification which is termed as \emph{presentation attack} or \emph{spoofing}. Spoofing has two main forms. One is physical access (PA) including direct imitation and replay. The other is logical access (LA) including speech synthesis and voice conversion. This paper focuses on the spoofing effects of text-to-speech (TTS) synthesis on SV systems. TTS has been investigated for decades. Early methods use unit selection \cite{hunt1996unit}, formant synthesis \cite{pinto1989formant}, and hidden Markov models (HMM) \cite{tokuda2000speech}, among others. Spoofing effects of synthetic speech from these traditional methods has been investigated in \cite{5444499,villalba2010speaker}. Recent years have witnessed the surge of deep learning speech synthesis models such as WaveNet \cite{oord2016wavenet} and Tacotron \cite{wang2017tacotron}; they can generate speech that is hard to be distinguished from real speech by listening. However, spoofing effects of synthetic speech from deep learning models have not been properly investigated \cite{sahidullah2019introduction}.\par
In \cite{tian2019blackbox}, a voice conversion (VC) system is proposed to spoof SV systems. It is trained using feedback from black-box SV systems for better spoofing effects. For TTS systems, a deep learning model based on GAN~\cite{cai2018attacking} is proposed to spoof an SV system by synthesizing mel-spectrograms. While promising results are reported, the spoofed SV system takes mel-spectrograms as input instead of time-domain signals. This is impractical in real life.
In \cite{7953088}, another GAN system is proposed to design a TTS system, which incorporates an anti-spoofing system as the discriminator. This system, however, is only evaluated on the feature distributions of the synthetic speech with and without GAN training; no evaluation is performed on spoofing SV systems.

In this paper, we propose a Wasserstein GAN-based multi-speaker TTS system based on an existing TTS \cite{tachibana2018efficiently} architecture to spoof SV systems\footnote{Source code at \url{https://github.com/MingruiYuan/SpoofSV}}. \zd{This system consists of two sub-models. The first model synthesizes a time-downsampled mel-spectrogram from text input using speaker embeddings of the target identity. The second model then converts the mel-specrogram to a linear-frequency spectrogram and finally to the time domain using the Griffin-Lim algorithm \cite{1164317}. We perform adversarial training for each sub-model.} Experiments are conducted on spoofing two state-of-the-art SV systems (i-vectors and Google GE2E) in a black-box condition, and results show a high spoof rate of the proposed TTS system. 

Our contributions are the following: 1) We proposed a multi-speaker TTS spoofing system using Wasserstein GAN training; 2) Comprehensive spoofing experiments showed a high spoof rate on two state-of-the-art SV systems in the black-box condition; 3) Our experiments also uncovered threats of TTS spoofing to anti-spoofing SV systems when their model structures are not kept confidential.    


\section{Text-to-speech Model}
\label{sec:tts}
Our proposed TTS model follows the two-stage process in \cite{tachibana2018efficiently}. In the first stage we use a Text2Mel network to convert the input text into a time-downsampled mel-spectrogram ``spoken'' by the target speaker. In the second stage we use a Spectrogram Super-resolution Network (SSRN) to convert the time-downsampled mel-spectrogram into the linear-frequency spectrogram. The Text2Mel model works in an online fashion: it processes acoustic features frame by frame. Previously generated frames of the features are fed back as input to Text2Mel to generate the next frame.




\subsection{Text2Mel Network}
\textbf{Text2Mel} consists of a text encoder (TEnc), an audio and speaker encoder (ASEnc) and an audio decoder (ADec). TEnc takes text embeddings as inputs. The text embeddings are obtained by mapping each character through a trainable lookup table. 
These embeddings are then processed by subsequent layers of TEnc to obtain output tensor $\mathbf{K, V}$. ASEnc has two input branches. One accepts the time-downsampled mel-spectrogram of previously generated audio frames and the other accepts the speaker embedding of the target speaker extracted by the Deep Speaker model \cite{DBLP:journals/corr/LiMJLZLCKZ17}. The two branches are added together and then processed by subsequent layers to obtain an output tensor $\mathbf{Q}$. 

Attention mechanism is employed to align the text input and the generated mel-spectrogram. This is implemented through a trainable attention matrix $\mathbf{A}$ with size $N \times T$, where $N$ is the total number of characters of the text input and $T$ is the total number of frames of the to-be-generated mel-spectrogram. $A_{nt}$ is the probability of the $t^{th}$ frame of the mel-spectrogram being generated from the $n^{th}$ character of the input text. As the alignment between text and its speech utterance is monotonic, during generation, we do not allow the alignment path to move backward in either dimension. In addition, we do not allow the path to skip 2 or more positions, leaving a valid step size of 0, 1 or 2 in both dimensions. This ensures a roughly continuous alignment path but also allows speed changes in the synthesized speech.\par 
Finally, ADec takes a concatenated tensor $[\mathbf{VA};\mathbf{Q}]$ as input and predicts a new frame of the time-downsampled melspectrogram in each time step, which is then appended to the generated spectrogram and fed to ASEnc in the next time step.\par

\begin{figure}[t]
    \centering
    \includegraphics[width=1.0\linewidth]{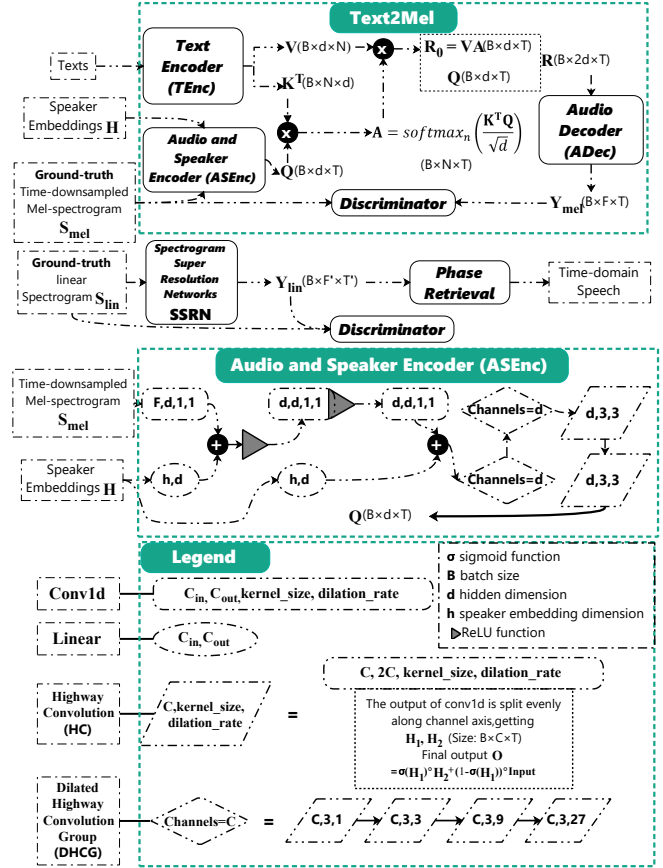}
    \caption{The proposed text-to-speech model. Architecture of ASEnc is illustrated, while that of TEnc, ADec and SSRN follows the same structure as that in \cite{tachibana2018efficiently}.}
    \label{fig:TTSmodel}
\end{figure}

In all modules of Text2Mel, 1D dilated convolutional layers are used to model short and long contextual information. Highway convolutional layers are applied according to highway networks \cite{srivastava2015highway} to improve training efficiency. Detailed structure of ASEnc is shown in Figure \ref{fig:TTSmodel}, while that of TEnc, ADec and SSRN follow the same structure as that in \cite{tachibana2018efficiently}.


\subsection{Spectrogram Super-resolution Network (SSRN)}
\textbf{SSRN} converts the time-downsampled mel-spectrogram from Text2Mel to the linear-frequency spectrogram. It uses transpose convolutional layers and a series of 1D dilated convolutional layers to achieve this super resolution along both time and frequency axes. Finally, Griffin-Lim algorithm \cite{1164317} is used to estimate the phase spectrogram to obtain the time-domain waveform of the generated speech.

\subsection{Training}
Training of Text2Mel and SSRN is performed separately, as shown in Figure \ref{fig:TTSmodel}. Training of Text2Mel requires pairs of text and time-downsampled mel-spectrograms,
while training of SSRN only requires pairs of time-downsampled mel-spectrograms and linear-frequency spectrograms.
As both networks only consist of 1D dilated convolutional layers, sequential models are avoided and all frames of each time-downsampled mel-spectrogram can be reconstructed at the same time with teacher-forcing in the training stage. This is the key to better training efficiency. The reconstruction loss functions for Text2Mel and SSRN are formulated as
\begin{small}
\begin{equation}
\begin{aligned}
  L_{Recon}^{Text2Mel} &=\mathbb{E}_{ft}[|\mathbf{Y_{mel}-S_{mel}}|]\\&+\mathbb{E}_{ft}[-\mathbf{S_{mel}}log\mathbf{Y_{mel}}-(1-\mathbf{S_{mel}})log(1-\mathbf{Y_{mel}})]\\&+L_{attention},
\end{aligned}
\end{equation}
\begin{equation}
\begin{aligned}
  L_{Recon}^{SSRN} &=\mathbb{E}_{ft}[|\mathbf{Y_{lin}-S_{lin}}|]\\&+\mathbb{E}_{ft}[-\mathbf{S_{lin}}log\mathbf{Y_{lin}}-(1-\mathbf{S_{lin}})log(1-\mathbf{Y_{lin}})],
\end{aligned}
\end{equation}\\
\end{small}
where $\mathbf{Y}$ denotes the reconstructed (mel-)spectrogram and $\mathbf{S}$ denotes the corresponding ground-truth. Both models use the $L_1$ and cross-entropy losses to assess the reconstruction quality. For Text2Mel, it also includes an attention loss term
$L_{attention} = \mathbb{E}_{nt}[\mathbf{A}\odot \mathbf{W}]$,
where the weight matrix $W_{nt} = 1-e^{-(\frac{n}{N}-\frac{t}{T})^2}$ shows a high weight off diagonal. This term penalizes the attention matrix $\mathbf{A}$ if it contains significant energy off diagonal. The rationale is that the alignment between text and its speech utterance is usually along the diagonal, assuming a stable speaking speed. 

When training the two sub-models we also incorporate discriminators that are trained to discriminate real and synthetic (mel-)spectrograms. The discriminators consist of 1D convolutional layers, highway convolutional layers, and 1D (adaptive) average pooling layers. Model details are omitted due to space limit but can be found in the open source code. We use Wasserstein GAN with gradient penalty (WGAN-GP)~\cite{NIPS2017_7159} because it achieves better results than the vanilla GAN in our experiments. Therefore, the output of each discriminator is a confidence value; A (mel-)spectrogram with a higher value is more likely to be real. The final loss function for Text2Mel and SSRN becomes: \cite{saito2017statistical}
\begin{small}
\begin{equation}
    L = L_{Recon} + \frac{\mathbb{E} {L_{Recon}}}{\mathbb{E} {L_{GAN}}}L_{GAN},
\end{equation}\\
\end{small}
where $L_{Recon}$ is the reconstruction loss function of each sub-model in Eqs. (1) and (2), while $L_{GAN}$ is the loss from the discriminator. The two parts in the loss function are normalized by their averages in each batch to have the same weight.\par

\section{Experiments}
\subsection{Training Spoofing Models}
We use the entire VCTK-corpus \cite{https://doi.org/10.7488/ds/1994} to train our TTS model. This corpus contains 108 valid English speakers (p315 is eliminated for the absence of texts) and each speaker has around 400 utterances ($\sim$0.5 hour). All audio files are downsampled from 48 kHz to 22.05 kHz and all texts are converted to lower case. 
We use STFT with a hanning window of size 1024 and hop size 256 to calculate the spectrogram. For the time-downsampled mel-spectrogram, we use 80 mel filterbanks, and select 1 frame out of every 4 frames. The optimizer is Adam \cite{kingma2014adam} with $\alpha=2e^{-4},\beta_1=0.5,\beta_2=0.9$ and batch size is 16. For every update of the generator, the discriminator is updated for 5 times. We set the gradient penalty coefficient $\lambda=10$. Layer normalization \cite{ba2016layer} is applied before each activation function as we found it useful in the experiments. We select three models trained for different number of iterations to perform experiments: $\mathbf{M_1}$ (Text2Mel) 500k-(SSRN) 300k. $\mathbf{M_2}$ 700k-500k. $\mathbf{M_3}$ 1000k-800k.


\subsection{Spoofing Effects on SV Systems}
We choose two state-of-the-art SV systems to spoof: \emph{i-vectors} \cite{dehak2010front} provided by \textit{Kaldi} and an open source implementation~\cite{ge2eurl} of Google's deep learning system \emph{GE2E}~\cite{wan2018generalized}.
We also use the VCTK corpus to train the SV systems. We split speakers in the corpus into training and test sets with three different schemes: $\mathbf{S_1}$ (Train) 42-(Test) 66. $\mathbf{S_2}$ 60-48. $\mathbf{S_3}$ 88-20. 
We use the default settings of \textit{Kaldi's aishell} example and the GE2E github repository to train the SV systems.

To investigate spoofing effects, we create a mixed set containing 50\% real and 50\% synthetic utterances to perform speaker verification. We randomly select 3 real utterances of each test speaker for enrollment. We then randomly choose another 20 real utterances and 20 synthetic utterances of each speaker for verification. The synthetic utterances are synthesized by the proposed TTS system on Harvard Sentences.
\begin{table}[t]
    \centering
    \begin{tabular}{cc|cc|cc}
    \toprule
    & &\multicolumn{2}{c|}{\textbf{i-vectors}}&\multicolumn{2}{c}{\textbf{GE2E}}\\
    & & SR & EER & SR & EER \\\midrule
     \multirow{3}{*}{$M_1$} & $S_1$ & 42.12\% & 1.97\% & 57.90\% & 19.35\% \\
         & $S_2$ & 42.40\% & 1.88\% & 77.36\% & 18.64\% \\
         & $S_3$ & 26\% & 1.5\% & 60.69\% & 18.57\% \\\hline
     \multirow{3}{*}{$M_2$} & $S_1$ & 70.15\% & 2.42\% & 62.39\% & 19.06\%  \\
         & $S_2$ & 72.08\% & 2.08\% & \textbf{80.20\%} & 18.49\% \\
         & $S_3$ & 41\% & 0.5\% & 65.07\% & 18.73\% \\\hline
     \multirow{3}{*}{$M_3$} & $S_1$ & \textbf{74.47\%} & 2.27\% & 69.80\% & 19.88\% \\
         & $S_2$ & 66.98\% & 1.67\% & 72.48\% & 19.68\% \\
         & $S_3$ & 57\% & 1.5\% & 69.16\% & 17.19\% \\\midrule
     \multicolumn{2}{c|}{Average} & 54.69\% & 1.75\% & 68.34\% & 18.74\% \\\bottomrule
    \end{tabular}
    \caption{Spoofing effects on speaker verification systems using three trained models ($M_1$, $M_2$, $M_3$) of the proposed TTS system with three data split schemes ($S_1$, $S_2$, $S_3$).}
    \label{tab:spoofres}
\end{table}

We propose \emph{spoof rate} (SR) to quantify the spoofing effects. It is defined as the percentage of synthetic speech utterances that are accepted by the SV system as their claimed identities. Apparently, SR is affected by the threshold tuning of SV systems. In this experiment, we tune the SV systems to achieve equal error rate (EER) on real speech utterances in the test set. In Table \ref{tab:spoofres}, we report SR of all the three models in the three train-test split schemes. We also report EER on real utterances as a control measure of the performance of the SV systems. From Table \ref{tab:spoofres}, we can see that all of the three models trained with the three data split schemes achieved a high spoof rate on both SV systems. The average spoof rate is 54.69\% for i-vectors and 68.34\% for GE2E. This shows the significant vulnerability of both SV systems under the attack of our TTS system. The EER on real utterances of i-vectors is below 2.5\% in all settings, showing that the SV system is well trained. The EER of GE2E is much higher, \zd{suggesting that} it is not well trained on our limited dataset. \zd{In fact, according to the GE2E's paper~\cite{wan2018generalized}, 18K speakers are used to train the model. Our training set, however, has less than 100 speakers for all the three data split schemes.} \zd{It is possible that a better trained GE2E model could be more robust to our TTS attack, and more investigations are needed to draw this conclusion.} 

\begin{figure}[t]
    \centering
    \includegraphics[width=0.9\linewidth]{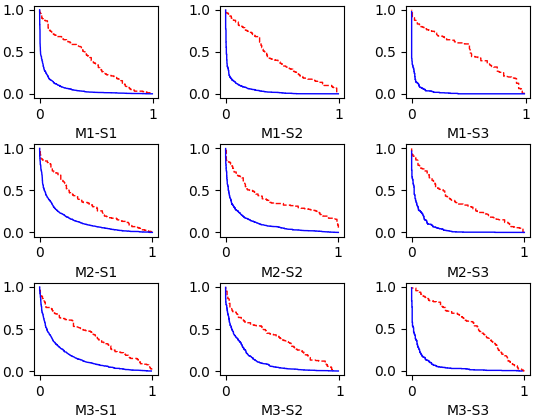}
    \caption{Spoof rate (horizontal axis) vs. false rejection rate (FRR) (vertical axis) of the three trained models ($M_1$, $M_2$, $M_3$) on three data split schemes ($S_1$, $S_2$, $S_3$). SV systems: i-vectors (blue solid line), GE2E (red dash line).}
    \label{fig:curve}
\end{figure}

In Figure \ref{fig:curve}, we vary the threshold of SV systems and plot the curve of spoof rate versus \emph{false rejection rate (FRR)}, where a false rejection is defined as the rejection of a real speech utterance that indeed belongs to the claimed identity. An ideal SV system that is robust to spoofing would show a monotonically decreasing curve very close to the origin. Curves in Figure \ref{fig:curve}, however, have a certain distance to the origin for all models and data split schemes. Take the M3-S3 i-vectors curve as an example, to lower the SR below 10\%, the FRR would be as high as 15\%, which would not be acceptable in practice. This again shows vulnerability of i-vectors and GE2E under the attack of our proposed TTS system.

\subsection{Spoofing Effects on Anti-spoofing Systems}
We further evaluate the proposed TTS model by spoofing anti-spoofing systems. Here, the anti-spoofing systems are binary classifiers that discriminate real from synthetic speech. We perform this evaluation in two conditions. In the \emph{blackbox} condition, the anti-spoofing system is treated as a blackbox and its model structure is not revealed to the TTS system. In the \emph{whitebox condition}, the model structure of the anti-spoofing system is revealed.
The anti-spoofing systems make two types of errors: 1) false acceptance of synthetic speech, and 2) false rejection of real speech. \zd{As the test set contains equal amount of real and synthetic utterances,} we report the equal error rate (EER) as the evaluation measure.


For the blackbox condition, we choose the \zd{ASVspoof2019 provided} GMM-based anti-spoofing system. It takes Linear Frequency Cepstral Coefficients (LFCC) as input features, and is trained on the logical access (LA) part of the ASVspoof2019 dataset~\cite{todisco2019asvspoof}. We compose two test sets, each of which is a mix of real speech utterances (50\%) and synthetic utterances (50\%). The difference is on the synthetic utterances. For the first set $\mathbf{TS_{proposed}}$, they are synthesized by the proposed TTS model. For the second set $\mathbf{TS_{others}}$, they are \zd{downloaded from \href{https://google.github.io/tacotron/}{https://google.github.io/tacotron/}} and are synthesized using other high-quality TTS models.
The resulted EER is 0.47\% on $\mathbf{TS_{proposed}}$ and 3.74\% on $\mathbf{TS_{others}}$. These low values shows the difficulty of spoofing anti-spoofing systems in the blackbox condition.\par

For the whitebox condition, we choose two variants $\mathbf{V_1}$ and $\mathbf{V_2}$ of the discriminator that we use in the GAN training of our model as the anti-spoofing system. Each variant has a similar structure to the original discriminator, \zd{with differences on the removal of an average pooling layer and the insertion of a convolutional layer, respectively.} We use test set $\mathbf{TS_{proposed}}$ to spoof $\mathbf{V_1}$ and $\mathbf{V_2}$. The resulted EER is 42.56\% for $\mathbf{V_1}$ and 36.21\% for $\mathbf{V_2}$. These high EER values show that both anti-spoofing systems fail to discriminate synthetic from real speech. This suggests that anti-spoofing systems, when their structures are disclosed, can be very vulnerable to TTS spoofing attacks.



\section{Conclusion and Future Work}
\zd{This paper proposed a deep multi-speaker TTS model for spoofing SV systems. GAN training was employed to train the two sub-models of the system. Experiments on spoofing state-of-the-art SV systems revealed their significant vulnerability under the attack of the proposed TTS system. Experiments on anti-spoofing systems also revealed their vulnerability if their model structures are disclosed. For future work, we plan to use reinforcement learning to improve the spoofing capability on blackbox SV systems. We also plan to design stronger anti-spoofing systems to defend TTS attack.}


\small
\bibliographystyle{IEEEbib}
\bibliography{bib/Deep-Speaker,bib/IVECTOR,bib/ge2e,bib/unitselection,bib/formant,bib/hmm,bib/wavenet,bib/gmmubm-sv,bib/GAN,bib/xvector,bib/dvector,bib/poveyemb,bib/security,bib/zaragoza,bib/JFA,bib/vad-overview,bib/DetectionHMM,bib/f01,bib/f02,bib/ASVspoof2015,bib/ASVspoof2017,bib/ASVspoof2019,bib/attackSV,bib/deceive,bib/VCattack,bib/TTSModel,bib/Griffinlim,bib/highway,bib/LN,bib/TTSGAN,bib/WGANGP,bib/Adam,bib/ge2eurl,bib/VCTK,bib/tacotron,bib/tacotronwavenet,bib/wavernn,bib/semi,bib/multilingual}

\end{document}